# Apparent giant dielectric constants, dielectric relaxation, and ac-conductivity of hexagonal perovskites $La_{1.2}Sr_{2.7}BO_{7.33}$ ($B$ = Ru, Ir)

P. Lunkenheimer[a,*], T. Götzfried[b], R. Fichtl[a], S. Weber[a], T. Rudolf[a], A. Loidl[a], A. Reller[b], and S.G. Ebbinghaus[b]

[a]*Experimental Physics V, Center for Electronic Correlations and Magnetism, University of Augsburg, D-86135 Augsburg, Germany*
[b]*Solid State Chemistry, University of Augsburg, D-86135 Augsburg, Germany*

We present a thorough dielectric investigation of the hexagonal perovskites $La_{1.2}Sr_{2.7}IrO_{7.33}$ and $La_{1.2}Sr_{2.7}RuO_{7.33}$ in a broad frequency and temperature range, supplemented by additional infrared measurements. The occurrence of giant dielectric constants up to $10^5$ is revealed to be due to electrode polarization. Aside of dc and ac conductivity contributions, we detect two intrinsic relaxation processes that can be ascribed to ionic hopping between different off-center positions. In both materials we find evidence for charge transport via hopping of localized charge carriers. In the infrared region, three phonon bands are detected, followed by several electronic excitations. In addition, these materials provide further examples for the occurrence of a superlinear power law in the broadband ac conductivity, which recently was proposed to be a universal feature of all disordered matter.

## 1. Introduction

During the last two decades, perovskite-related oxides have attracted broad attention in solid state physics, as they reveal a variety of extraordinary magnetic and electronic properties like high-$T_c$ superconductivity [1], colossal magnetoresistance [2], or multiferroic behavior [3]. Besides their interesting physical properties, these materials show a remarkable compositional flexibility. Especially, in addition to the two archetypical packing structures, i.e. the cubic and the hexagonal perovskite, there is the possibility of forming a huge variety of layered structures consisting of perovskite slabs intergrown by layers of other structural types. The recently synthesized title compounds are formed by alternative stacking of hexagonal perovskite ($A_2BO_6$) and $A'_2O_{1+\delta}$ layers (A = La/Sr; A' = Sr; B = Ru, Ir) [4,5]. Detailed structural investigations revealed the presence of both, oxygen and peroxide ions in the $A'_2O_{1+\delta}$ layers. These materials therefore belong to the few examples, in which complex ions are incorporated into a perovskite-like lattice. Moreover, these ions, occupying large cavities formed by six triangular $AO_6$ prisms, are not located in the center of these cavities, but can assume six different degenerate off-center positions. Finally, also the A'-ions in the $A'_2O_{1+\delta}$ layers can occupy three multiple positions [4,5].

Ions in off-center positions are prone to interesting dielectric phenomena as dipolar relaxation or polar ordering, paramount examples being the dielectrically active Ti-ions in $SrTiO_3$ or $BaTiO_3$. In addition, the strong substitutional disorder of $La_{1.2}Sr_{2.7}MO_{7.33}$ should give rise to disorder-induced localization of charge carriers and thus the ac conductivity of these materials is expected to show the typical signature of hopping transport. Finally, this work is part of a systematic investigation of the broadband ac conductivity behavior of disordered semiconductors, aiming at the broadening of the experimental data base concerning the universal occurrence of a superlinear power law (SLPL) [6]. Therefore we have performed a thorough investigation of the dielectric and ac transport properties of the title compounds, covering a broad frequency range of more than ten decades. The results are complemented by infrared spectra, giving a detailed picture of the relaxational and electronic properties. Aside of the detection of giant values of the dielectric constant ascribed to interfacial polarization effects, we find two significant intrinsic relaxation processes in both materials. As expected, they also exhibit pronounced ac conductivity contributions due to hopping charge transport. In addition, clear indications for a SLPL of the frequency-dependent conductivity were found.

## 2. Experimental Details

Details on sample preparation can be found in Refs. [4,5]. Polycrystalline pellets were cold-pressed and sintered in air for 10 h at 1050 K. For most of the dielectric measurements, opposite sides of the pellets were coated with silver-paint, thus forming a parallel-plate capacitor. To clarify the influence of electrode polarization effects, additional measurements with sputtered gold contacts (thickness 100 nm) were performed.

Measurements of the dielectric properties at frequencies between 0.1 Hz and 3 GHz and temperatures 20 – 550 K were achieved by combining different experimental setups [7]. Up to 3 MHz, a frequency-response analyzer (Novocontrol α-analyzer) was used. The samples were cooled down to about 20 K in a closed-cycle-refrigerator while heating up to 550 K was achieved in a $N_2$-gas cryostat (Novocontrol quatro). In the radio to microwave regime, a reflectometric technique employing an impedance analyzer (Agilent 4991A, 1 MHz – 3 GHz) was applied, the sample shorting the inner and outer conductor at the end of a home-made coaxial line. For cooling, the end of the coaxial line was connected to the cold head of a closed-cycle refrigerator, allowing measurements down to 25 K. A self-designed furnace was employed for measurements up to 550 K.

[*]Corresponding author. Tel.: +49-821-5983603; fax: Tel.: +49-821-5983649.
*E-mail address:* Peter.Lunkenheimer@Physik.Uni-Augsburg.de

Investigations in the infrared regime were performed on highly polished pellets. The reflectivity was determined using two Fourier-transform spectrometers (Bruker IFS 113v and IFS66v/S). A spectral range from 50 cm$^{-1}$ ($\approx$ 1.5 THz) to 25000 cm$^{-1}$ ($\approx$ 750 THz) was covered utilizing a suitable set of sources, beamsplitters, windows, and detectors. Reference measurements were performed with a gold mirror at $\nu < 12000$ cm$^{-1}$ and an aluminum mirror at higher frequencies. Spectra of $\sigma'$ and $\varepsilon'$ were calculated from the measured reflectivity and the phase shift, the latter being determined via Kramers-Kronig (KK) transformation. For this purpose, at low frequencies a constant extrapolation and at high frequencies a $\nu^{-1}$ power-law extrapolation followed by a $\nu^{-4}$ law at $\nu > 10^6$ cm$^{-1}$ were used.

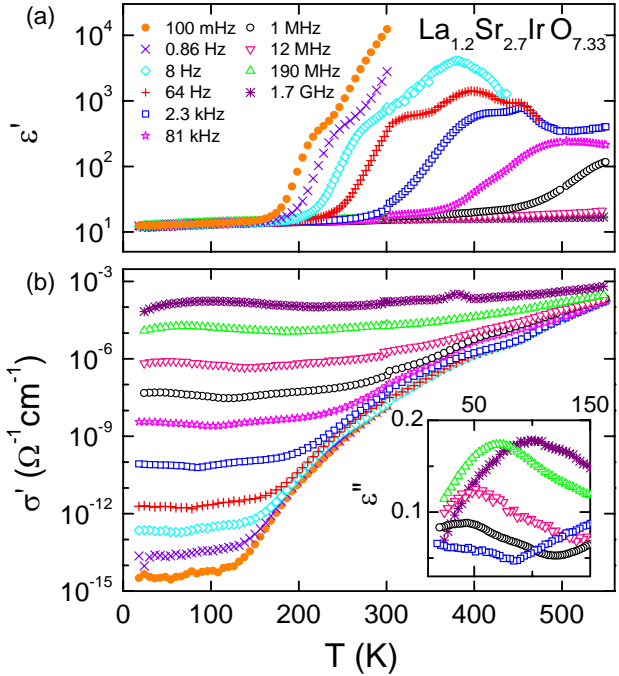

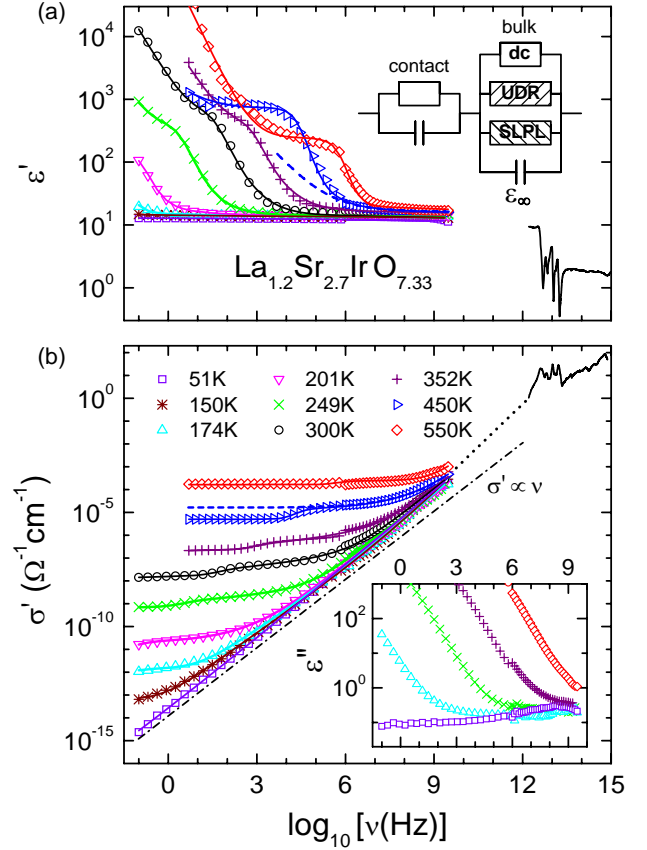

Fig. 1. Temperature dependent dielectric constant (a) and conductivity (b) of La$_{1.2}$Sr$_{2.7}$IrO$_{7.33}$ at various frequencies. The inset shows the dielectric loss in the region of the low-temperature relaxation.

Fig. 2. Dielectric constant (a) and conductivity (b) of La$_{1.2}$Sr$_{2.7}$IrO$_{7.33}$ at various temperatures. The infrared results, collected at room temperature, are shown as solid lines. The dash-dotted line in (b) indicates a linear increase; the dotted line interpolates the GHz and infrared results by an SLPL. The inset in (b) shows the dielectric loss at selected temperatures. In (a), the equivalent circuit used to fit the experimental data is shown. The resulting fits, performed simultaneously for $\varepsilon'$ and $\sigma'$, are shown as solid lines. The dashed lines in (a) and (b) demonstrate the intrinsic response at 450 K.

## 3. Results

Figure 1(a) shows the temperature-dependent dielectric constant $\varepsilon'$ of La$_{1.2}$Sr$_{2.7}$IrO$_{7.33}$ for various frequencies. With increasing temperature, $\varepsilon'(T)$ exhibits a strong steplike increase to values around $10^3$, which for the lower frequencies is superimposed by a further increase up to $10^4$. The steps shift to higher temperatures with increasing frequency thus showing the typical signature of relaxational behavior. The relaxational character of $\varepsilon'$ is also revealed by its frequency dependence, given in Fig. 2(a), where characteristic relaxation steps are observed. The further increase at low frequencies (corresponding to that at high temperatures in Fig. 1(a)) is either due to ac conductivity or a second relaxation process. At high frequencies, $\varepsilon'(\nu)$ levels off at a value of about 14. This frequency-independent value is usually denoted as $\varepsilon_\infty$ in dielectric spectroscopy and ascribed to the ionic and electronic polarizability. As can be seen from the inclusion of the infrared results in Fig. 2(a), it corresponds to the low-frequency limit of the infrared results, where a number of phononic and electronic excitations are observed, which will be discussed below.

The conductivity $\sigma'(T)$ is shown in Fig. 1(b). For the lower frequencies, its temperature dependence, which is identical to that of the dielectric loss $\varepsilon'' \propto \sigma'/\nu$, shows a strong increase with temperature at $T > 130$ K. At high temperatures and low frequencies, this contribution is nearly frequency-independent and thus can be assumed to reflect the dc conductivity. Coming from high temperatures, the $\sigma'(T)$ curves for different frequencies branch off from this dc-like contribution, exhibiting a weak temperature dependence only, typical for ac conductivity due to hopping transport of localized charge carriers [8]. As it is characteristic for ac conduction, in this region $\sigma'(\nu)$ increases significantly with frequency, which is demonstrated in the frequency-dependent plot of Fig. 2(b). At the lowest temperatures and/or highest frequencies, this increase is slightly steeper than linear as becomes obvious by comparison to the dash-dotted line indicating a linear increase. This behavior corresponds to an SLPL in $\sigma'(\nu)$ or a nearly constant loss in $\varepsilon''(\nu)$ (inset of Fig. 2(b)). At low frequencies and the higher temperatures, $\sigma'(\nu)$ becomes nearly frequency independent, showing the signature of dc conductivity, except for a small steplike variation, e.g. located at about $10^5$ Hz for 450 K. This feature, which corresponds to a peak in $\varepsilon''(\nu)$ (not shown), is Kramers-Kronig related to the strong relaxation steps in $\varepsilon'(\nu)$. The corresponding contribution in the temperature dependence (Fig 1(b)) shows up as a shoulder superimposed to the increasing dc-like part of $\sigma'(\nu)$, barely visible due to the high density of data points in this region.

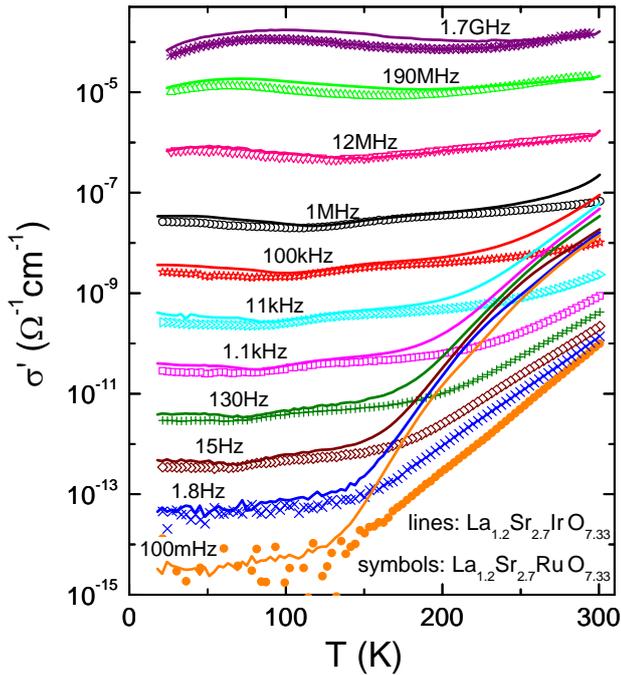

Fig. 3. Comparison of the temperature-dependent conductivity of $La_{1.2}Sr_{2.7}IrO_{7.33}$ and $La_{1.2}Sr_{2.7}RuO_{7.33}$ for various frequencies.

In Fig. 1(b), two further weak relaxational contributions are observed, superimposed to the ac conductivity. They show up as weak peaks or shoulders (e.g. located at about 50 K and 200 K for the 12 MHz curve), which shift to higher temperatures with increasing frequency. The inset of Fig. 1 shows the dielectric loss in the region of the more pronounced low-temperature relaxation, indeed revealing significant loss peaks. This feature is best seen in the temperature dependence and is only weakly visible in $\varepsilon''(\nu)$ (inset of Fig. 2(b)), e.g., at $10^8$ Hz for the 450 K curve.

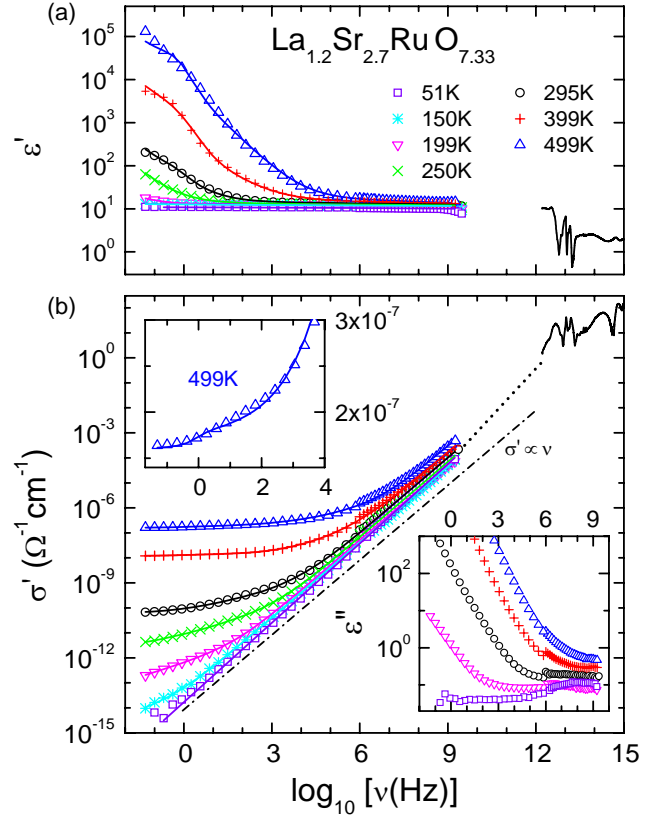

Fig. 4. Dielectric constant (a) and conductivity (b) of $La_{1.2}Sr_{2.7}RuO_{7.33}$ at various temperatures. The infrared results, collected at room temperature, are shown as solid lines. The dash-dotted line in (b) indicates a linear increase; the dotted line interpolates the GHz and infrared results by an SLPL. The solid lines show fits with the equivalent circuit indicated in Fig. 2, performed simultaneously for $\varepsilon'$ and $\sigma'$. The upper inset in (b) gives a magnified view of the low-frequency behavior at 499 K. The lower inset shows the dielectric loss at selected temperatures.

It is clear that a variety of different superimposed contributions lead to a rather complex dielectric behavior in $La_{1.2}Sr_{2.7}IrO_{7.33}$. In $La_{1.2}Sr_{2.7}RuO_{7.33}$ the situation is very similar, however, with some quantitative differences as revealed by Figs. 3 and 4. The ac conductivity and the small low-temperature relaxations of the ruthenate are very similar to those of the iridate (Fig. 3). However, the dc-like part of the conductivity (roughly corresponding to the 100 mHz curves at $T > 150$ K) is much lower for the ruthenate (Fig. 3). For example, the dc conductivity at room temperature for $La_{1.2}Sr_{2.7}RuO_{7.33}$ can be estimated as about $10^{-10}$ $\Omega^{-1}cm^{-1}$ while that of the iridate reaches at least $10^{-8}$ $\Omega^{-1}cm^{-1}$. In addition, the main relaxation steps in $\varepsilon'(\nu)$ are shifted to much lower frequencies as becomes obvious by comparing Figs. 4(a) and 2(a). At first glance, there seems to be no steplike increase in $\sigma'(T)$ corresponding to the relaxation feature in $\varepsilon'(T)$, in contrast to the finding in $La_{1.2}Sr_{2.7}IrO_{7.33}$. However, the upper inset of Fig. 4(b) reveals that there is indeed such a contribution, but with a rather weak amplitude only.

### 3. Discussion

In both investigated materials, the dominating feature in the temperature and frequency dependence of the dielectric loss at sub-Hz to GHz frequencies is a strong relaxational contribution leading to very high dielectric constants at low frequencies and/or high temperatures. In recent years, an increasing number of publications on a variety of different materials have appeared that report a qualitatively similar dielectric behavior, namely the occurrence of very large dielectric constants in conjunction with a strong relaxation mode (e.g., [9,10,11,12,13]). New high-$\varepsilon$ materials that would avoid the disadvantages of the currently employed ferroelectric materials, namely their strong temperature and field dependence, are highly sought-after for the construction of more effective capacitive elements in electronics. However, in [14] it was argued that most, if not all, detections of "giant" or "colossal" dielectric constants reported so far most likely are due to interface polarization effects. For example, depletion layers, arising from the formation of Schottky diodes at the metallic contacts of semiconducting samples, or so-called "internal barrier layer capacitors" (IBLCs) [15], which may be formed, e.g., by grain boundaries, can give rise to Maxwell-Wagner type relaxations and apparently very high dielectric constants.

To check for such effects in the current investigation, the silver paint was removed from the $La_{1.2}Sr_{2.7}IrO_{7.33}$ sample in an ultrasonic bath after the measurements and sputtered gold contacts were applied. If contact effects dominate the dielectric response, sputtered contacts usually lead to the higher apparent dielectric constants [14]. This can be explained by a more effective formation of the Schottky barriers because the very small metal clusters applied during sputtering will lead to a larger area of direct metal-semiconductor contact than for the relatively large particles ($\geq \mu m$) suspended in the silver paint. In contrast, due to the similar work function of silver and gold and the fact that their resistivities are negligibly small compared to that of the sample material, the utilization of different metals has no influence on the results. In Fig. 5, the dielectric constant and conductivity spectra obtained with both types of contacts are compared. Obviously, there is a very strong effect of contact type on the dominating relaxation: For the sputtered contacts, the low-frequency limiting dielectric constant considerably increases and the relaxation features in $\varepsilon'$ and $\sigma'$ are shifted to lower frequencies by 1-2 decades. The amplitude of the observed conductivity step is significantly smaller for the sputtered contacts, however, $\sigma'(\nu)$ at high frequencies is nearly identical for both measurements. Also the $\varepsilon'(\nu)$ curves for different contact types approach similar values at high frequencies.

This behavior is typical for a non-intrinsic Maxwell-Wagner relaxation due to electrode polarization [14]. The contact depletion-layer can usually be well approximated by an equivalent circuit with a parallel RC circuit in series to the bulk contribution as indicated in Fig. 2(a). With increasing frequency, the contact capacitor becomes shorted and the intrinsic response is detected. Therefore, at high frequencies the curves obtained with both types of contacts approach each other. At low frequencies, $\varepsilon'$ is dominated by the extrinsic contact capacitance, which usually is much higher than the intrinsic bulk capacitance. As described above, due to the better "wetting" achieved by sputtered contacts, their contact capacitance is higher than for silver paint contacts. The fact that the relaxation time of the equivalent circuit can be expected to increase with increasing contact capacitance explains the shift of the relaxation features to lower frequencies for the sputtered sample. The higher limiting low-frequency conductivity observed for the sputtered contacts in Fig. 5(b) indicates that the contact resistance is slightly smaller for this type of contact preparation.

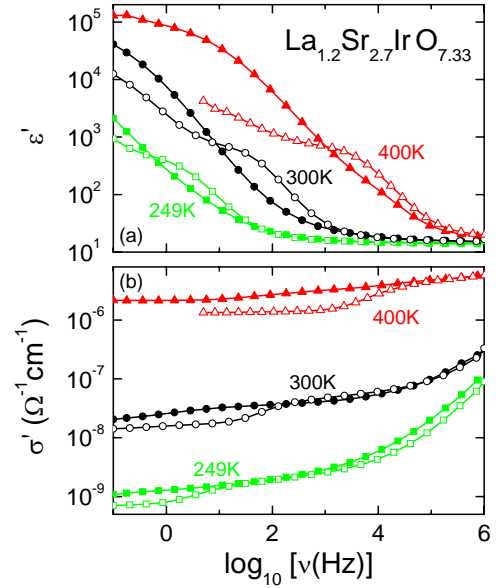

Fig. 5. Dielectric constant (a) and conductivity (b) of $La_{1.2}Sr_{2.7}IrO_{7.33}$ using silver paint (open symbols) and sputtered gold contacts (closed symbols). The lines connect the data points.

To further analyze the experimental results and obtain information on the intrinsic properties, least-square fits of the frequency-dependent data with the equivalent circuit shown in Fig. 1(a) were performed [14,16]. In this circuit, the bulk contribution is modeled by a resistor for the dc-conductivity and a capacitor for the limiting high-frequency dielectric constant $\varepsilon_\infty$. In addition, a further parallel element is added to cover the SLPL, $\sigma' \propto \nu^n$ including its contribution to the dielectric constant, $\varepsilon' \propto -\nu^{n-1}$. In [6], the SLPL was shown to be a widespread phenomenon in many different disordered materials (e.g., [11,16,17,18]). However, similar to the findings in other transition-metal oxides [6,16,17,18], the dc conductivity and the SLPL alone are not sufficient to properly fit the rather smooth transition region between the low-frequency dc plateau and the high-frequency SLPL seen in Figs. 2(b) and 4(b). At low temperatures (150 K for $La_{1.2}Sr_{2.7}IrO_{7.33}$ and 150 K and 199 K for $La_{1.2}Sr_{2.7}RuO_{7.33}$), another, sublinear power law is revealed. Such a power law corresponds to the so-called "Universal Dielectric Response" (UDR), which was demonstrated by Jonscher [19] to be a universal feature in the dielectric response of disordered matter. For conducting materials, it is commonly ascribed to the hopping of charge carriers subjected to disorder-induced localization [8]. Such a localization is a common phenomenon in amorphous materials, but can also arise from the substitutional disorder in highly doped semiconductors. To take into account this contribution, another element (sometimes termed "constant phase element" in literature) was added to the equivalent circuit, assuming $\sigma' \propto \nu^s$ and the Kramers-Kronig related contribution to the dielectric constant, $\varepsilon' \propto \nu^{s-1}$ [19]. We refrained from including another element to account for the small intrinsic relaxation features discussed above, as these represent only relatively weak contributions to the loss (insets of Figs. 2(b) and 4(b)) and are barely visible in $\sigma'(\nu)$.

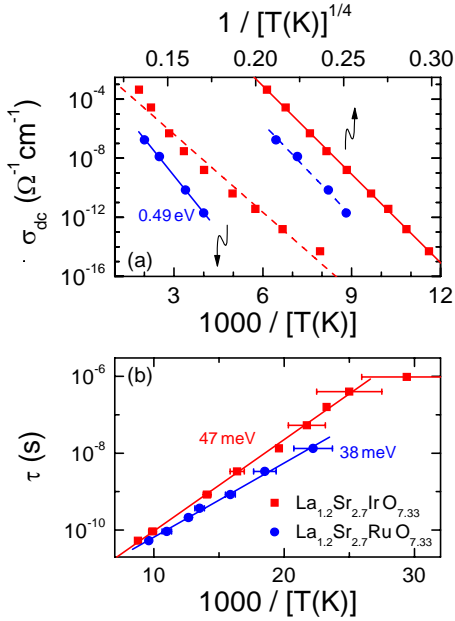

Fig. 6. (a) Temperature-dependent dc conductivity of $La_{1.2}Sr_{2.7}IrO_{7.33}$ (squares) and $La_{1.2}Sr_{2.7}RuO_{7.33}$ (circles) obtained from the fits of the frequency-dependent data (Figs. 2 and 4). The data are shown in an Arrhenius representation (lower scale) and in a representation leading to a linearization for the VRH model (upper scale). The lines are linear fit curves. (b) Relaxation times of the intrinsic relaxation at $T < 100$ K (Figs. 1(b) and 3) shown in an Arrhenius representation. The data points were determined by reading off the peak positions in the temperature dependent plots. The lines are linear fits, demonstrating thermally activated behavior.

The solid lines shown in Figs. 2 and 4 are fits using this equivalent circuit, performed simultaneously for $\varepsilon'(\nu)$ and $\sigma'(\nu)$. Obviously, the experimental data can be well described in this way. For $T = 450$ K, the dashed lines in Fig. 2 demonstrate the intrinsic bulk contribution, again proving that the step in $\sigma'(\nu)$ and the strong relaxation mode in $\varepsilon'(\nu)$ are due to the contacts, described by the parallel RC-circuit. The intrinsic $\varepsilon'(\nu)$ strongly increases with decreasing frequency, according to the $\nu^{s-1}$ UDR power law. This contribution also leads to the further increase in the measured $\varepsilon'(\nu)$ observed at low frequencies (e.g. at $\nu < 10^4$ Hz for 550 K) in Fig. 2(a). This notion was checked by simulating curves omitting the UDR element, which leads to the vanishing of the low-frequency increase. Concerning the parameters resulting from the fits, the different regions as dc conduction, UDR, and SLPL are not sufficiently separated to allow for a precise determination of all parameters. In particular, this holds for the exponent parameters $s$ and $n$. However, from Figs. 2 and 4 and especially from the low-temperature loss data shown in the insets, $n$ can be estimated to be about 1.03 for $La_{1.2}Sr_{2.7}IrO_{7.33}$ and 1.02 for $La_{1.2}Sr_{2.7}RuO_{7.33}$, both values being consistent with the fit results.

The intrinsic dc conductivity $\sigma_{dc}$, determined from the fits, is shown in Fig. 6(a). Due to the lower conductivity of the ruthenium compound (cf. Fig. 3), $\sigma_{dc}$ could only be determined at $T \geq 250$ K. The data are shown in an Arrhenius plot (lower scale in Fig. 6(a)) and in a representation that leads to a linearization of the curves for Mott`s variable range hopping (VRH) model [20]. The VRH model considers the phonon-assisted tunneling of charge carriers that are subjected to disorder-induced localization. It predicts a behavior $\sigma_{dc} \sim \exp[-(T_0/T)^{1/4}]$, where $T_0$ is proportional to $\alpha^3/N(E_F)$, with $\alpha$ the inverse of the localization length and $N(E_F)$ the density of states at the Fermi level [20]. For $La_{1.2}Sr_{2.7}RuO_{7.33}$, the Arrhenius representation seems to provide a slightly better description of the experimental data. However, due to the rather restricted temperature range a clear statement is not possible (Fig. 6(a)). For $La_{1.2}Sr_{2.7}IrO_{7.33}$, on the other hand, the experimental data significantly deviate from a simple thermally-activated Arrhenius behavior and the VRH seems to be the dominating dc transport process. For the ac conductivity, the VRH predicts a UDR power law [8,20], i.e. its detection in the dc conductivity is consistent with our findings in the ac conductivity. VRH was also observed in various other transition metal oxides (see, e.g., [13,16,17,18,21]), with values of $T_0$ ranging between $1\times10^7$ K [21] and $8\times10^8$ K [13]. For $La_{1.2}Sr_{2.7}IrO_{7.33}$ we find a somewhat higher value of $T_0 = 5.4\times10^9$ K. It should be noted that the exponent $\gamma = 1/4$ of the VRH expression for $\sigma_{dc}(T)$ is altered to 1/2 or 1/3 if the charge transport becomes restricted to one or two dimensions, respectively. As the crystal structure of the title compounds is highly anisotropic, such a scenario seems possible. Indeed, the data for the iridate can also be described using $\gamma = 1/3$ with nearly equal quality (not shown), and it is not possible to reach a definite conclusion concerning the dimensionality of the charge transport.

As mentioned above, two intrinsic relaxation processes are detected in both investigated materials at low temperatures (cf. Figs. 1(b) and 3). The one located at lower temperatures is of sufficient magnitude to allow for a quantitative evaluation. To obtain information on the corresponding relaxation time, we evaluated the position of the loss peaks in the temperature-dependent data (e.g., inset of Fig. 1(b)), where this feature can be best resolved. The relaxation time at the peak temperature then can be calculated via $\tau = 1/(2\pi\nu)$ with $\nu$ the measuring frequency. The resulting $\tau(T)$ is shown in Fig. 6(b) in an Arrhenius representation. For both materials, the relaxation times are of similar magnitude, varying over many decades, which mirrors a continuous slowing down of relaxational motion with decreasing temperature. $\tau(T)$ can be well approximated by a thermally activated behavior with energy barriers of 47 meV and 38 meV for the Ir and Ru compound, respectively. Considering the structural details of the investigated materials [4,5], it seems reasonable to ascribe the microscopic origin of the two observed relaxations to the hopping of ions between different degenerate off-center positions. Specifically, the oxygen and/or peroxide ions can hop between six positions at the site termed "O2" in [5] and the Sr ion can take three different positions at the "Sr2" site. However, presently it is not possible to decide, which of these modes causes which of the two observed intrinsic relaxations.

Finally, in Fig. 7 the infrared results are shown for $La_{1.2}Sr_{2.7}IrO_{7.33}$ and $La_{1.2}Sr_{2.7}RuO_{7.33}$. With values around 10, the low-frequency limit of $\varepsilon'$ nicely matches $\varepsilon_\infty$ observed at high frequencies in the dielectric measurements (cf. Figs. 2(a) and 4(a)). This proves the intrinsic nature of this value and the absence of any additional excitations in the frequency region between 3 GHz and 1.5 THz ($9.5 < \log_{10}\nu < 12.2$), which is not covered by our experiments. In the region between 3 THz and 30 THz ($12.5 < \log_{10}\nu < 13.5$), resonance-like features show up, typical for phononic excitations. As was also found for other perovskite-related materials (e.g., [17,18,22]), there are three main contributions, with peaks in $\sigma'(\nu)$ at about 5 THz, 10 THz, and 16 THz ($\approx$ 170 cm$^{-1}$, 330 cm$^{-1}$, and 530 cm$^{-1}$, respectively). In cubic perovskites, $ABO_3$, the three infrared-active phonon excitations are ascribed to external, bending, and stretching modes. They are due to the vibration of the $BO_6$ octahedra against the A ions, the bending of the B-O bond-angle, and the variation of the B-O bond length, respectively. In $La_{1.2}Sr_{2.7}IrO_{7.33}$ and $La_{1.2}Sr_{2.7}RuO_{7.33}$ the two high-frequency modes are very similar for both materials and resemble those observed in other perovskite-related compounds. This is consistent with the fact that the bending and stretching modes essentially are local modes of the $BO_6$ octahedra. In contrast, the first peak at about 5 THz in both materials is highly broadened and in $La_{1.2}Sr_{2.7}IrO_{7.33}$ is composed of at least two separate peaks. This strong deviation from the findings in simple perovskites is reasonable as this mode is determined by more global lattice vibrations. The observed broadening most likely is due to the superposition of several heavily damped phonon modes. The phonon damping may result from the disordered arrangement of the A' cations and/or the oxygen ions in the crystal structure or from a hybridization of the phonon excitations with relaxational modes. The significant difference in the infrared-active density of states for the two compounds remains unexplained. For a detailed analysis, factor group analysis needs to be performed, and spectra with higher resolution need to be collected, which is beyond the scope of the present work.

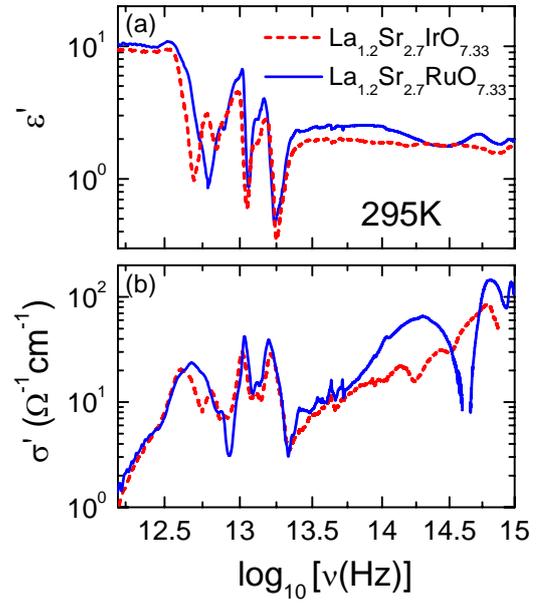

Fig. 7. Dielectric constant (a) and conductivity (b) of $La_{1.2}Sr_{2.7}IrO_{7.33}$ and $La_{1.2}Sr_{2.7}RuO_{7.33}$ in the infrared region at room temperature.

It should be noted that a broad loss peak at low infrared frequencies, accompanied by a relaxation-like behavior of the dielectric constant as observed in the present case, is a common feature of glasses and supercooled liquids and termed "boson peak" in these materials (see, e.g., [23]). It corresponds to modes in excess of the purely Debye-derived vibrational density of states but its detailed microscopic origin still is controversially discussed. Recently it was shown [24] that a boson-peak like excitation also occurs in the so-called plastic crystals, where the centers of mass of the molecules form a regular crystalline lattice but the dipolar molecules are disordered with respect to their orientational degrees of freedom [25]. As detailed above, in the materials investigated in the present work, oxygen/peroxide and Sr ions can assume different degenerate off-center positions. This situation is analogous to dipoles with orientational degrees of freedom and thus these materials could be regarded as a special type of plastic crystals. Therefore one may speculate that the unusually broad first excitation observed in the infrared spectra of $La_{1.2}Sr_{2.7}IrO_{7.33}$ and $La_{1.2}Sr_{2.7}RuO_{7.33}$ (Fig. 7) and the so-far mysterious boson peak in glass formers and plastic crystals have a common origin.

At frequencies beyond the phonon modes, significant differences in the infrared response of both materials show up. Especially, $La_{1.2}Sr_{2.7}RuO_{7.33}$ exhibits two pronounced peaks in $\sigma'(\nu)$ at about 200 and 650 THz while only much weaker excitations are detected in the iridate. This is the region of electronic or polaronic excitations and it is interesting that the 0.49 eV activation energy determined from $\sigma_{dc}(T)$ of $La_{1.2}Sr_{2.7}RuO_{7.33}$ (Fig. 6(a)) corresponds to a frequency of about $10^{14}$ Hz, which is close to the first excitation in both compounds (Fig. 7(b)). Finally, we want to remark that the strong peak at 200 THz in the ruthenate resembles a polaronic excitation as observed, e.g., in perovskite manganites [26]. To corroborate this conjecture, infrared measurements at different temperatures need to be performed to check for the characteristic temperature dependence of polaronic excitations.

## 3. Summary and Conclusions

We have reported a comprehensive investigation of the dielectric and ac-conductivity properties of the perovskite-related compounds $La_{1.2}Sr_{2.7}IrO_{7.33}$ and $La_{1.2}Sr_{2.7}RuO_{7.33}$ in a broad frequency and temperature range, complemented by room-temperature infrared measurements. In both materials a strong relaxational mode and the appearance of extremely large dielectric constants up to $10^5$ at low frequencies and high temperatures were detected, arising from electrode polarization and ac conductivity effects. Charge transport in these materials is dominated by hopping charge carriers that are localized due to substitutional disorder, the dc conductivity of the iridate being significantly higher than that of the ruthenate. In addition to the UDR behavior of the ac conductivity, we find evidence for an SLPL at high frequencies and low temperatures as promoted recently to be a universal feature of all disordered matter [6]. Two intrinsic relaxation processes could be detected in both compounds, which tentatively are ascribed to the hopping of oxygen ions ($O^{2-}$ and/or peroxide ions $(O_2)^{2-}$) and strontium ions between multiple off-center positions. In the infrared region, three main phonon bands, typical for perovskite-related oxides, are seen, the two high-frequency ones being due to oscillation modes of the $BO_6$ octahedra. The first band is significantly broadened, mirroring the deviations from the ideal perovskite structure.

## Acknowledgements

This work was supported by the Deutsche Forschungsgemeinschaft via the Sonderforschungsbereich 484 and by the BMBF via VDI/EKM.